\documentclass[pre,aps,showpacs,amsmath,twocolumn,a4paper,superscriptaddress]{revtex4-1}%
\usepackage{graphicx}
\usepackage{dcolumn}
\usepackage{bm}
\usepackage{amsmath}
\usepackage{amssymb}
\usepackage{color}
\makeatother

\usepackage{color}

\begin{document}

\title{
When do microscopic assumptions determine \\
the outcome in evolutionary game dynamics?
}

\author{Bin Wu}
\email{bin.wu@evolbio.mpg.de}
\affiliation{Department of Evolutionary Theory,\\ Max Planck Institute for Evolutionary Biology, Pl\"{o}n, Germany}
\author{Benedikt Bauer}
\email{bauer@evolbio.mpg.de}
\affiliation{Department of Evolutionary Theory,\\ Max Planck Institute for Evolutionary Biology, Pl\"{o}n, Germany}
\author{Tobias Galla}
\email{tobias.galla@manchester.ac.uk}
\affiliation{Theoretical Physics, School of Physics and Astronomy, The University of 
Manchester, Manchester M13 9PL, United Kingdom}
\author{Arne Traulsen}
\email{traulsen@evolbio.mpg.de}
\affiliation{Department of Evolutionary Theory,\\ Max Planck Institute for Evolutionary Biology, Pl\"{o}n, Germany}

\date{\today}

\begin{abstract}
The modelling of evolutionary game dynamics in finite populations requires microscopic processes that determine how strategies spread. 
The exact details of these processes are often chosen without much further consideration. 
Different types of microscopic models, including in particular fitness-based selection rules and imitation-based dynamics, are often used as if they were interchangeable. 
We challenge this view and investigate how robust these choices on the micro-level really are. 
Focusing on a key macroscopic observable, the probability for a single mutant to take over a population of wild-type individuals, we show that there is a unique pair of a fitness-based process and an imitation process leading to identical outcomes for arbitrary games and for all intensities of selection. 
This highlights the perils of making arbitrary choices at the micro-level without regard of the consequences at the macro-level.
\end{abstract}

\maketitle

\label{sec:M}

Evolutionary game theory is a powerful framework to model social and biological evolution when the success of an individual depends on the presence or absence of other strategies \cite{nowak:Science:2004,szabo:PR:2007,roca:PLR:2009,perc:BioSys:2010}. 
In this context, the payoff from a game between individuals is translated into reproductive fitness. 
Methods from statistical physics have been applied extensively since the field moved from mostly deterministic models based on rate equations to stochastic individual-based models \cite{helbing:PA:1993b,traulsen:PRL:2005,antal:PRL:2006,ohtsuki:PRL:2007,van-segbroeck:PRL:2009,black:PRL:2012}.  
These more sophisticated models use a microscopic process as a starting point to determine how successful strategies spread. 
Tools and ideas from statistical physics are key to making the connection between the assumptions on the micro-scale, and effective descriptions on the macro-scale. 
Two classes of microscopic processes have been used extensively: 
(i) Fitness-based processes in which an individual chosen proportional to fitness reproduces and the offspring replaces a randomly chosen individual \cite{nowak:Nature:2004}; 
(ii) Imitation-based processes in which a pair of individuals is chosen, and where subsequently one of these individuals may adopt the strategy of the other.
This adaptation occurs with a probability that depends on the payoff of both individuals, such that better players are more likely to be imitated than those who do worse \cite{blume:GEB:1993,szabo:PRE:1998}.  
The payoff-to-fitness mapping used in the context of fitness-based processes can be interpreted as a transformation between Malthusian fitness (the growth rate, which can be negative) and Wrightian fitness (the average number of offspring in the next generation, a non-negative quantity) \cite{buerger:book:2000,Wu:EaE:2013}. 

In types of processes, the relative influence of the game is controlled by an external parameter,
the so-called intensity of selection $\beta$.  
This parameter has strong parallels to the inverse temperature in statistical mechanics \cite{traulsen:PRE:2006b}. 
In populations of size $N$ the dynamics is dominated by the evolutionary game for strong selection, $\beta N \gg 1$, with demographic noise only affecting the outcome weakly. For weak selection, $\beta N \ll 1$, the dynamics is largely stochastic, with only a small influence of the game on the evolution of the system. The outcome of evolutionary game dynamics thus depends on the interplay between selection and noise, both changing with the relative abundance of the types of individuals in the population. 
In well-mixed populations and on some special networks (e.g., on a ring) 

the evolutionary dynamics between two types of strategies, wild-type and mutant, can be described by simple birth-death processes. In such processes the state of the system is characterized by the number of mutants alone. A quantity that is of particular interest in evolutionary biology is the probability of fixation, which is the likelihood that a mutant type takes over the entire population \cite{karlin:book:1975,ewens:book:1979}. It is the basis of the definition of evolutionary stability in finite populations \cite{nowak:Nature:2004}. It also features in the leading-order term of a small-mutation expansion of the stationary distribution which serves as a powerful analytical method when multiple strategies are present in the population \cite{fudenberg:JET:2006,wu:JMB:2012}.

The choice of a fitness-based processes versus an imitation-based processes is typically not further justified in the literature \cite{fu:PRE:2008,sigmund:Nature:2010}. Often the type of model employed is chosen arbitrarily. 
 This is frequently no cause for concern as many results do not seem to depend on the particular choice of the microscopic process. 
 In particular, a wide class of microscopic processes leads to similar results under weak selection \cite{lessard:JMB:2007,lessard:DGAA:2011b}. 
 However, this equivalence is only partial, and in some cases the outcome on the macro-scale can crucially depend on the specific choices made at the microscopic level \cite{wu:PLoSCB:2013}. 
 Here, we show the choice of an fitness-based versus a imitation-based process is restricted to a unique pair if we require that, for any {\em arbitrary} game, the two processes lead to identical fixation probabilities for {\em all} intensities of selection $\beta$. This indicates that the choice of the microscopic process can make a difference even in unstructured populations. The outcome is independent of the underlying microscopic process only for weak selection (i.e., in the high-temperature regime) and for constant selection.

\label{sec:R}
We consider well-mixed populations with fixed size $N$. 
Each individual can be of one of two types, $A$ and $B$. The state of the population is thus characterized by the number $i$ of individuals of type $A$. The interaction between the two types of individuals is described by the functions $\pi_A^i$ and $\pi_B^i$. These indicate the expected payoff for two types in a population in state $i$. The interaction can be thought of as a two-player matrix game \cite{hofbauer:book:1998}, but we keep the formalism general to include games played between an arbitrary number of players \cite{gokhale:DGAA:2014,pena:JTB:2014}.

A discrete-time birth-death process on the set of states $i=0,\ldots, N$ 
is characterized by the transition probabilities $T^{i\pm}$ that the system moves to state $i\pm 1$ in the next step, when it is currently in state $i$. 
With probability $1-T^{i+}-T^{i-}$, the system remains in state $i$. 
We restrict ourselves to processes for which $T^{i\pm}>0$ for all $i=1,\ldots,N-1$, and in which the two states $i=0$ and $i=N$ are absorbing, i.e. $T^{0+}=T^{N-}=0$. The population can never escape from homogenous states.  In biology this corresponds to the absence of mutation, where extinct types cannot be re-introduced.

We will now characterize fitness-based processes and imitation-based processes in more detail. For a given game, i.e.\ for payoff functions $\pi_A^i$ and $\pi_B^i$, a fitness-based process assumes that at each time step an individual is selected for reproduction with a probability proportional to its fitness. 
This individual produces one identical offspring which replaces a randomly chosen individual in the population. Consequently, the transition probabilities are of the form
\begin{equation}
\label{eq:transitionMoran}
T_{F}^{i+}=\frac{i}{N} \frac{f_A^i}{\langle f \rangle_i} \frac{N-i}{N},
\quad
T_{F}^{i-}=\frac{N-i}{N} \frac{f_B^i}{\langle f \rangle_i} \frac{i}{N}.
\end{equation}
The subscript `$F$' indicates a fitness-based process. We have assumed that the payoffs $\pi_{A}^i$ and $\pi_B^i$ translate into reproductive fitness via a mapping $f_A^i=f(\beta\pi_A^i)$ and $f_B^i=f(\beta\pi_B^i)$, where $\beta>0$ is the intensity of selection and where $f'(x)> 0$ for all $x$, indicating that fitness increases with payoff. The quantity $\langle f\rangle_i$ is the average fitness of an individual in the population, i.e., $\langle f\rangle_i=\left(if_A^i+(N-i)f_B^i\right)/N$. The transition rates in Eq.~(\ref{eq:transitionMoran}) are then fully specified by the underlying game and by the payoff-to-fitness mapping $f$.

In an imitation process, one focal individual and a role model are chosen at random at each time step. 
The payoff difference between the two individuals determines the probability that the focal individual adopts the strategy of the role model. 
Specifically, for a focal individual of type $A$ and a role model of type $B$, this probability is $g[\beta(\pi_B^i-\pi_A^i)]$, where $\beta>0$ is again the intensity of selection. If the focal individual is of type $B$ and the role model of type $A$ this probability is $g[\beta(\pi_A^i-\pi_B^i)]$. 
The derivative $g'(x)$ of the imitation function $g(x)$ must be positive to ensure it is more likely to adopt successful strategies. For a given game and a given adaptation function $g$ this leads to a birth-death process with the transition 
probabilities
\begin{equation}
\label{eq:transtion Imitation}
T_I^{i\pm}=\frac{i(N-i)}{N^2}g\left[\pm\beta(\pi_A(i)-\pi_B(i))\right].
\end{equation}
The the subscript `I' indicates an imitation process.
 
For both classes of processes, and for any game, the dynamics will eventually reach one of the two absorbing states: Either the mutant goes extinct (absorption at $i=0$), or it reaches fixation ($i=N$). The so-called fixation probability, $\phi$, measures how likely it is that a single mutant takes over the entire population, i.e. it is the probability for the system to end up in $i=N$, if initialised at $i=1$. For general birth-death processes this probability is given by
\cite{nowak:Nature:2004,karlin:book:1975,ewens:book:1979}
\begin{align}
\label{eq:fixationprob}
 \phi = \left(\sum_{k=0}^{N-1} \prod_{i=1}^{k} \frac{T^{i-}}{T^{i+}}\right)^{-1}.
\end{align}
Our arguments hold also for the case of fixation from an arbitrary number of mutants,
but we focus on the biologically most relevant scenario of a single mutant. 
The central result of our paper concerns the following question: 
For what choices of the payoff-to-fitness mapping, $f$, and of the imitation function, $g$, do the resulting fitness-based and imitation-based processes have the same fixation probability, $\phi_F=\phi_I$, for {\em arbitrary} games and intensities of selection?
In other words,
if we require that the two processes are equivalent in fixation for any game and any selection intensity, how do we need to choose these two processes?

We note that $T_F^{i-}/T_F^{i+}=f(\beta\pi_B^i)/f(\beta\pi_A^i)$ for the fitness-based process, and $T_I^{i-}/T_I^{i+}=g[\beta(\pi_B^i-\pi_A^i]/g[(\beta\pi_A^i-\pi_B^i)]$ for the imitation process. If the functions $f$ and $g$ fulfill 
\begin{equation}
\label{eq:fg}
\frac{f(x)}{f(y)}=\frac{g(x-y)}{g(y-x)},
\end{equation}
for all $x, y$, we have $T_F^{i-}/T_F^{i+}=T_I^{i-}/T_I^{i+}$ for all $i$. Using Eq.\ (\ref{eq:fixationprob}) this leads to equal fixation probabilities for all games and any selection intensity. 
Thus, Eq.\ \eqref{eq:fg} is sufficient.

We now show that Eq.~\eqref{eq:fg} is also necessary. 
To this end we note that the functions $f$ and $g$ must be such that the equality of fixation probability holds for {\em all} games, so in particular for games with constant $\pi_A^i = \pi_A$ and $\pi_B^i = \pi_B$. 
For such games the ratios $\gamma_F = T_F^{i-}/T_F^{i+}$ and $\gamma_I = T_I^{i-}/T_I^{i+}$ are independent of $i$. The equality of fixation probabilities is then equivalent to $p(\gamma_F)=p(\gamma_I)$, where $p(\gamma)=\sum_{\ell=0}^{N-1} \gamma^\ell$, cf.\ Eq.\ (\ref{eq:fixationprob}). 
The polynomial $p(\gamma)$ is strictly increasing for positive arguments. 
Considering that both $\gamma_F$ and $\gamma_I$ are positive,
$p(\gamma_F)=p(\gamma_I)$ implies $\gamma_F=\gamma_I$. 
The constants $\pi_A$ and $\pi_B$ can be chosen arbitrarily, as the selection intensity $\beta$. The fact that we require $\gamma_F=\gamma_I$ leads to the conclusion that $f$ and $g$ must fulfil Eq.\ \eqref{eq:fg}.
Eq.\ \eqref{eq:fg} is thus necessary if we require identity of fixation times for {\em all} possible games. 
We stress that it may well be possible to construct a game and a pair of functions $f$ and $g$, which are not of the above form, such that the fixation probabilities of the two resulting processes coincide for this particular game. 
However, unless $f$ and $g$ fulfil Eq. (\ref{eq:fg}) the identity of fixation probabilities will not hold for arbitrary games, as our argument above shows. 

Eq.\ (\ref{eq:fg}) implies that the ratio $f(x)/f(y)$ has to be a function of the difference $x-y$ alone. Setting $y=x+\Delta x$ in $f(x)/f(y)=g(x-y)/g(y-x)$ and taking the limit $\Delta x\to 0$ leads to the differential equation
\begin{equation}
\label{eq:odefg}
\frac{ f'(x)}{f(x)} = 2\frac{g'(0)}{g(0)}.
\end{equation}
We note that this differential equation must hold for all $x$. It is a necessary condition for the equality of fixation probabilities for arbitrary games and arbitrary strength of selection, but it is not a sufficient condition by itself. A necessary and sufficient condition is given by Eq. (\ref{eq:fg}). 

\medskip

We observe that the condition of Eq. (\ref{eq:odefg}) can be relaxed if we limit the equality of fixation probabilities to the weak-selection approximation.
It corresponds to expanding the fixation probabilities to linear order in the selection intensity.
If we require that $f$ and $g$ lead to identical fixation probabilities only in the linear-order term in $\beta$ (but not necessarily to higher order) for any payoff functions $\pi_A^i$ and $\pi_B^i$ we obtain the condition 
\begin{equation}
\label{eq:univer-weak}
\frac{f'(0)}{f(0)}=2\frac{g'(0)}{g(0)}.
\end{equation}
  
This condition is far less restrictive than Eq. (\ref{eq:odefg}), and it is both necessary and sufficient to have identity of fixation times for all games up to linear order in $\beta$. This can be seen from existing results for weak selection \cite{wu:PRE:2010}. 
The only solution of the more restrictive condition, Eq.~\eqref{eq:odefg}, is
\begin{equation}
\label{eq:RLSH 1}
f(x)=f(0)\exp\left[2\frac{g'(0)}{g(0)}x\right].
\end{equation}
 
This implies that in order for the fixation probabilities of a fitness-based process to be identical to those of an imitation based process (to any order in the selection intensity), it is necessary that the payoff-to-fitness mapping $f(\beta\pi)$ is exponential in $x$, $f(x)=f(0)\exp(\lambda x)$, where $\lambda$ is an arbitrary positive constant.  The imitation function $g$ is at this point largely unconstrained, although one finds $g(x)/g(-x)=e^{\lambda x}$ by setting $y=0$ in Eq. (\ref{eq:fg}). With the additional assumption $g(x)+g(-x)=1$,  only a single possible imitation functions remains, the so-called Fermi function $g(x)=[1+e^{-\lambda x}]^{-1}$.

\begin{figure}[t!!]
\centering
\includegraphics[width=0.5\textwidth]{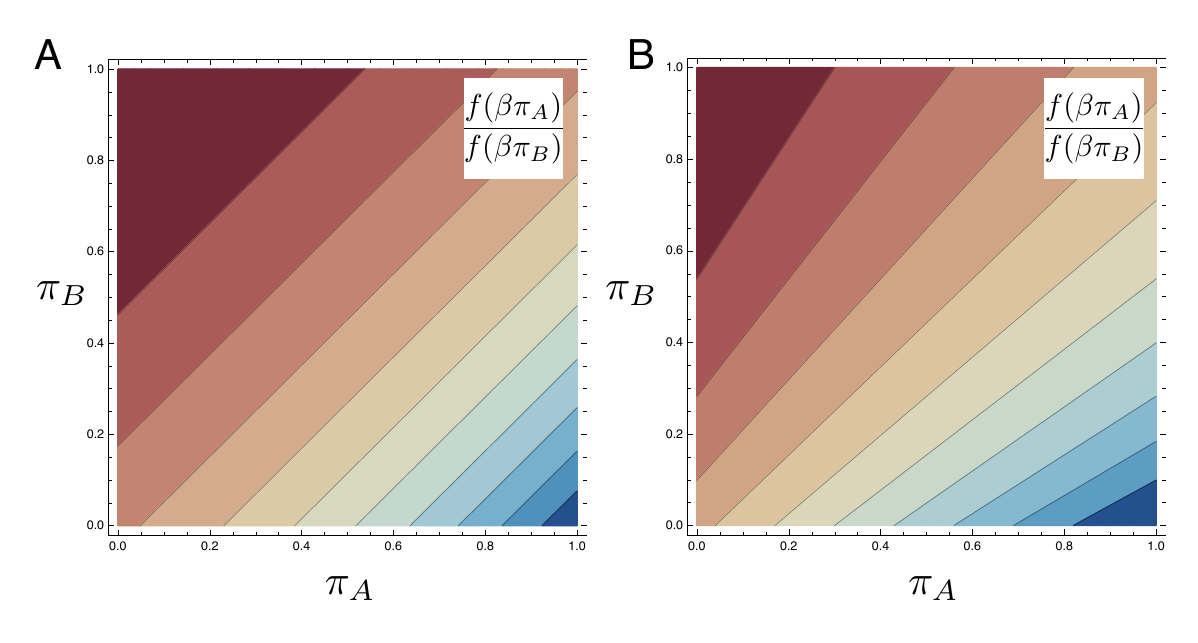}
\caption{Graphical representation of Eq.~\eqref{eq:fg}.
The contour plot depicts the ratio $f(\beta \pi_A)/f(\beta \pi_B)$. 
Panel A: For the exponential payoff-to-fitness mapping $f(x)=\exp(x)$, the ratio $f(\beta \pi_A)/f(\beta \pi_B)$
obviously depends on the difference $\pi_A-\pi_B$ only, such that the contour lines are diagonal in the $\pi_A-\pi_B$ plane. An identical picture is obtained for $g[\beta(\pi_A-\pi_B)]/g[\beta(\pi_B-\pi_A)]$ with $g(x) = [1+\exp(-x)]^{-1}$.  Panel B: For the affine linear payoff-to-fitness mapping $f(x)=1+x$, the ratio $f(\beta \pi_A)/f(\beta \pi_B)$ depends on $\pi_A$ and $\pi_B$ explicitly, not only on their difference. The contour lines are no longer parallel to each other. It is not possible to reconstruct an imitation function $g$ leading to equivalent fixation times for all games, and which depends on payoff difference only. For simplicity we have used $\beta=1$ in the figure.}
\label{fig:ill}
\end{figure}

We have thus shown that the assumption of equal fixation probabilities for {\em all} games together with the mild assumption $g(x)+g(-x)=1$ fully restricts the payoff-to-fitness mapping and the imitation function to $f(x)=f(0)e^{\lambda x}$ and $g(x)=1/(1+e^{-\lambda x})$. 
The only remaining free parameters are $f(0)$ and the constant $\lambda$. 
However, the choice of $f(0)$ is immaterial as $f(0)$ drops out in Eq.\ (\ref{eq:transitionMoran}). The constant $\lambda$ on the other hand can effectively be absorbed in the selection strength, $\beta$, so that, to all intents and purposes, our constraints fully specify the payoff-to-fitness mapping and the imitation function. Thus, this pair of processes is unique and, if chosen otherwise, the precise details of the microscopic model will affect the outcome of the model on the macroscopic level.  For example, the popular linear payoff-to-fitness mapping $f=1+\beta \pi$ has no corresponding imitation function which depends on payoff differences only and which leads to the same fixation probability for arbitrary games. This is illustrated in Fig.~\ref{fig:ill}.

The allowed set of imitation functions becomes broader if we relax the constraint and allow functions $g$ with $g(x)+g(-x)\neq 1$. We find that any imitation function of the form $g(x)=h(x)/(1+e^{-\lambda x})$ is permissible so long as the resulting $g(x)$ is increasing, takes values between $0$ and $1$ (such that it is a probability), 
and $h(x)$ is even (to ensure $f(x)/f(y)=g(x-y)/g(y-x)$). To show that such functions $h(x)$ exist we mention two arbitrary examples,
$h(x)=\exp[- \tfrac{1}{2}e^{-x^2} ]$ and $h(x)=\tfrac{2}{3} \pm \tfrac{1}{3 \cosh[x]}$.

We now consider more general imitation processes in which the imitation probability does not depend on payoff differences alone. Specifically, we allow imitation probabilities with which a focal individual with payoff $\pi_{\rm foc}$ imitates the strategy of a role model with payoff $\pi_{\rm rm}$ of the form $Q(\beta\pi_{\rm foc},\beta\pi_{\rm rm})$, i.e.\ $Q$ may depend on the payoffs of both individuals explicitly. The previous case is recovered as $Q(x,y)=g(y-x)$. To guarantee that the resulting imitation function $Q(x,y)$ is a probability, it has to take values between $0$ and $1$. In addition, we require  $\partial_x Q(x,y) < 0$, such that focal individuals with high payoff are less likely to adopt the strategies of others, and $\partial_y Q(x,y) > 0$, such that role models with higher payoff are more likely to be imitated than those with a low payoff.
In this more general case, a fitness-based process with payoff-to-fitness mapping $f$ has the same fixation probability of a single mutant as an imitation process if $f(x)/f(y)=Q(y,x)/Q(x,y)$ in analogy to Eq.~\eqref{eq:fg}.
Setting $y=x+\Delta x$ and taking the limit $\Delta x\to0$ leads to the necessary condition
\begin{align}
\label{Eq:ODE-general}
f'(x)=\Gamma(x) f(x), 
\end{align}
where $\Gamma(x) = Q(x,x)^{-1}\left.\left[\left(\partial_y-\partial_x\right) Q(x,y)\right]\right|_{y=x}$. 
From this, one obtains
\begin{equation}
\label{Eq:M+I Gener}
f(x)= f(0)\exp\left[\int_0^x\ \Gamma(z) d z\right].
\end{equation}
Again this condition is necessary, but not sufficient by itself to guarantee equal fixation probabilities under both processes for arbitrary games. 
Condition (\ref{Eq:M+I Gener}) admits payoff-to-fitness mappings $f(x)$ that are not exponential. The constraint that $f(x)$ must be exponential in $x$ derived under the more restrictive assumptions above is a specific consequence of the requirement that the imitation probability depends on payoff differences only.

For any given payoff-to-fitness mapping  $f(x)$ which is increasing and positive,
the 
function $Q(x,y)=f(y)/(f(x)+f(y))$ proposed in \cite{nowak:Nature:2004} is decreasing in $x$ and increasing in $y$ and takes values between $0$ and $1$.
In other words, it fulfills the constraints of an imitation function.
Thus, for any payoff-to-fitness mapping, $f(x)$, there is an imitation kernel $Q(x,y)$ leading to equal fixation probabilities for all games. Restricting the set of permissible kernels to those of the form $Q(x,y)= g[\psi(y)-\psi(x)]$ with $g(x)+g(-x)=1$ and where $\psi(x)$ is an increasing function fully specifies the imitation kernel. A short calculation shows that the imitation function $Q(x,y)=f(y)/[f(x)+f(y)]$ is then the only possible imitation choice leading to identical fixation probabilities for all games for a given payoff-to-fitness mapping.  For the exponential mapping, $f(x)=f(0)e^{\lambda x}$, this is the Fermi function $Q(x,y)=e^{\lambda x}/(e^{\lambda x}+e^{\lambda y})=1/[1+e^{-\lambda (x-y)}]$.
 
In addition to the fixation probability, the so-called gradient of selection is often used to investigate stochastic evolutionary game dynamics \citep{pinheiro:NJP:2012}. 
It is given by $T^{i+}-T^{i-}$ and represents the drift term of the corresponding Fokker-Planck equation in the diffusion approximation \cite{traulsen:PRE:2012,bladon:PRE:2010}. The qualitative features of the deterministic flow of a given game will generally not depend upon the choice of a fitness-based Moran process versus a pairwise comparison process. However, going beyond this qualitative level immediately reveals that a fitness-based process and an imitation-based process cannot have identical gradient of selection for all $i$ and all $x$ and $y$. In order for this to be the case we require
 \begin{equation}
\label{Equ:BothCases}
 \frac{f(\beta\pi_A^i) - f(\beta\pi_B^i)}{\langle f \rangle_i}  = g[\beta(\pi_A^i-\pi_B^i)]-g[\beta(\pi_B^i-\pi_A^i)].
 \end{equation}
However, the left-hand side explicitly depends on the state $i$, whereas the right-hand side depends on $i$ only via the payoffs $\pi_A^i$ and $\pi_B^i$. These can be chosen not to depend on $i$, so Eq.~(\ref{Equ:BothCases}) cannot be valid for arbitrary games.
In particular, such an equivalence is not found for the pair of processes that leads to identical fixation probabilities discussed above. 
Under weak selection, such an identity in the gradient of selection is often found. However the requirements for an equivalence of selection gradient in the weak selection approximation are generally quite different from those needed to result in an equivalence of fixation probabilities in the same limit.

\medskip

In summary we have challenged some of the key assumptions frequently made in modelling evolutionary dynamics. Fitness-based and imitation-based processes are often used as if these approaches were entirely exchangeable. This is appropriate -- to a certain extent -- when fitness is a positive constant as it is the case in many models of classical population genetics. The choice of the microscopic details of the process does however make a difference for the macroscopic outcome of frequency-dependent selection outside the regime of weak selection. As we have shown there are then strong restrictions on the choice of the imitation function and the payoff-to-fitness mapping if one requires that the fixation probabilities in the two classes of processes are identical for any intensity of selection. Additionally we find that it is impossible to construct two such processes that lead to the same gradient of selection and to identical fixation probabilities. These challenges are largely absent in population genetics, where selection is constant, and only arise in evolutionary game theory, where selection is frequency dependent. In evolutionary games on graphs a dependence on the microscopic details has been pointed out repeatedly \cite{szabo:PR:2007,roca:PLR:2009,perc:BioSys:2010}. 
It is noteworthy that these difficulties are already present in non-spatial well-mixed systems of the type that we have discussed. The complexity of a networked structure is therefore not a necessary component. Indeed, we would expect it to be much more challenging to construct two processes with identical outcomes on such more complicated geometries.

\begin{acknowledgments}
We thank Benjamin Werner for helpful discussions. 
\end{acknowledgments}


\begin{thebibliography}{32}%
\makeatletter
\providecommand \@ifxundefined [1]{%
 \@ifx{#1\undefined}
}%
\providecommand \@ifnum [1]{%
 \ifnum #1\expandafter \@firstoftwo
 \else \expandafter \@secondoftwo
 \fi
}%
\providecommand \@ifx [1]{%
 \ifx #1\expandafter \@firstoftwo
 \else \expandafter \@secondoftwo
 \fi
}%
\providecommand \natexlab [1]{#1}%
\providecommand \enquote  [1]{``#1''}%
\providecommand \bibnamefont  [1]{#1}%
\providecommand \bibfnamefont [1]{#1}%
\providecommand \citenamefont [1]{#1}%
\providecommand \href@noop [0]{\@secondoftwo}%
\providecommand \href [0]{\begingroup \@sanitize@url \@href}%
\providecommand \@href[1]{\@@startlink{#1}\@@href}%
\providecommand \@@href[1]{\endgroup#1\@@endlink}%
\providecommand \@sanitize@url [0]{\catcode `\\12\catcode `\$12\catcode
  `\&12\catcode `\#12\catcode `\^12\catcode `\_12\catcode `\%12\relax}%
\providecommand \@@startlink[1]{}%
\providecommand \@@endlink[0]{}%
\providecommand \url  [0]{\begingroup\@sanitize@url \@url }%
\providecommand \@url [1]{\endgroup\@href {#1}{\urlprefix }}%
\providecommand \urlprefix  [0]{URL }%
\providecommand \Eprint [0]{\href }%
\providecommand \doibase [0]{http://dx.doi.org/}%
\providecommand \selectlanguage [0]{\@gobble}%
\providecommand \bibinfo  [0]{\@secondoftwo}%
\providecommand \bibfield  [0]{\@secondoftwo}%
\providecommand \translation [1]{[#1]}%
\providecommand \BibitemOpen [0]{}%
\providecommand \bibitemStop [0]{}%
\providecommand \bibitemNoStop [0]{.\EOS\space}%
\providecommand \EOS [0]{\spacefactor3000\relax}%
\providecommand \BibitemShut  [1]{\csname bibitem#1\endcsname}%
\let\auto@bib@innerbib\@empty
\bibitem [{\citenamefont {Nowak}\ and\ \citenamefont
  {Sigmund}(2004)}]{nowak:Science:2004}%
  \BibitemOpen
  \bibfield  {author} {\bibinfo {author} {\bibfnamefont {M.~A.}\ \bibnamefont
  {Nowak}}\ and\ \bibinfo {author} {\bibfnamefont {K.}~\bibnamefont
  {Sigmund}},\ }\href@noop {} {\bibfield  {journal} {\bibinfo  {journal}
  {Science}\ }\textbf {\bibinfo {volume} {303}},\ \bibinfo {pages} {793}
  (\bibinfo {year} {2004})}\BibitemShut {NoStop}%
\bibitem [{\citenamefont {Szab{\'o}}\ and\ \citenamefont
  {F{\'a}th}(2007)}]{szabo:PR:2007}%
  \BibitemOpen
  \bibfield  {author} {\bibinfo {author} {\bibfnamefont {G.}~\bibnamefont
  {Szab{\'o}}}\ and\ \bibinfo {author} {\bibfnamefont {G.}~\bibnamefont
  {F{\'a}th}},\ }\href@noop {} {\bibfield  {journal} {\bibinfo  {journal}
  {Physics Reports}\ }\textbf {\bibinfo {volume} {446}},\ \bibinfo {pages} {97}
  (\bibinfo {year} {2007})}\BibitemShut {NoStop}%
\bibitem [{\citenamefont {Roca}\ \emph {et~al.}(2009)\citenamefont {Roca},
  \citenamefont {Cuesta},\ and\ \citenamefont {Sanchez}}]{roca:PLR:2009}%
  \BibitemOpen
  \bibfield  {author} {\bibinfo {author} {\bibfnamefont {C.~P.}\ \bibnamefont
  {Roca}}, \bibinfo {author} {\bibfnamefont {J.~A.}\ \bibnamefont {Cuesta}}, \
  and\ \bibinfo {author} {\bibfnamefont {A.}~\bibnamefont {Sanchez}},\
  }\href@noop {} {\bibfield  {journal} {\bibinfo  {journal} {Physics of Life
  Reviews}\ }\textbf {\bibinfo {volume} {6}} (\bibinfo {year}
  {2009})}\BibitemShut {NoStop}%
\bibitem [{\citenamefont {Perc}\ and\ \citenamefont
  {Szolnoki}(2010)}]{perc:BioSys:2010}%
  \BibitemOpen
  \bibfield  {author} {\bibinfo {author} {\bibfnamefont {M.}~\bibnamefont
  {Perc}}\ and\ \bibinfo {author} {\bibfnamefont {A.}~\bibnamefont
  {Szolnoki}},\ }\href@noop {} {\bibfield  {journal} {\bibinfo  {journal}
  {BioScience}\ }\textbf {\bibinfo {volume} {99}},\ \bibinfo {pages} {109}
  (\bibinfo {year} {2010})}\BibitemShut {NoStop}%
\bibitem [{\citenamefont {Helbing}(1993)}]{helbing:PA:1993b}%
  \BibitemOpen
  \bibfield  {author} {\bibinfo {author} {\bibfnamefont {D.}~\bibnamefont
  {Helbing}},\ }\href@noop {} {\bibfield  {journal} {\bibinfo  {journal}
  {Physica A}\ }\textbf {\bibinfo {volume} {196}},\ \bibinfo {pages} {546}
  (\bibinfo {year} {1993})}\BibitemShut {NoStop}%
\bibitem [{\citenamefont {Traulsen}\ \emph {et~al.}(2005)\citenamefont
  {Traulsen}, \citenamefont {Claussen},\ and\ \citenamefont
  {Hauert}}]{traulsen:PRL:2005}%
  \BibitemOpen
  \bibfield  {author} {\bibinfo {author} {\bibfnamefont {A.}~\bibnamefont
  {Traulsen}}, \bibinfo {author} {\bibfnamefont {J.~C.}\ \bibnamefont
  {Claussen}}, \ and\ \bibinfo {author} {\bibfnamefont {C.}~\bibnamefont
  {Hauert}},\ }\href@noop {} {\bibfield  {journal} {\bibinfo  {journal}
  {Physical Review Letters}\ }\textbf {\bibinfo {volume} {95}},\ \bibinfo
  {pages} {238701} (\bibinfo {year} {2005})}\BibitemShut {NoStop}%
\bibitem [{\citenamefont {Antal}\ \emph {et~al.}(2006)\citenamefont {Antal},
  \citenamefont {Redner},\ and\ \citenamefont {Sood}}]{antal:PRL:2006}%
  \BibitemOpen
  \bibfield  {author} {\bibinfo {author} {\bibfnamefont {T.}~\bibnamefont
  {Antal}}, \bibinfo {author} {\bibfnamefont {S.}~\bibnamefont {Redner}}, \
  and\ \bibinfo {author} {\bibfnamefont {V.}~\bibnamefont {Sood}},\ }\href@noop
  {} {\bibfield  {journal} {\bibinfo  {journal} {Physical Review Letters}\
  }\textbf {\bibinfo {volume} {96}},\ \bibinfo {pages} {188104} (\bibinfo
  {year} {2006})}\BibitemShut {NoStop}%
\bibitem [{\citenamefont {Ohtsuki}\ \emph {et~al.}(2007)\citenamefont
  {Ohtsuki}, \citenamefont {Nowak},\ and\ \citenamefont
  {Pacheco}}]{ohtsuki:PRL:2007}%
  \BibitemOpen
  \bibfield  {author} {\bibinfo {author} {\bibfnamefont {H.}~\bibnamefont
  {Ohtsuki}}, \bibinfo {author} {\bibfnamefont {M.~A.}\ \bibnamefont {Nowak}},
  \ and\ \bibinfo {author} {\bibfnamefont {J.~M.}\ \bibnamefont {Pacheco}},\
  }\href@noop {} {\bibfield  {journal} {\bibinfo  {journal} {Physical Review
  Letters}\ }\textbf {\bibinfo {volume} {98}},\ \bibinfo {pages} {108106}
  (\bibinfo {year} {2007})}\BibitemShut {NoStop}%
\bibitem [{\citenamefont {Van~Segbroeck}\ \emph {et~al.}(2009)\citenamefont
  {Van~Segbroeck}, \citenamefont {Santos}, \citenamefont {Lenaerts},\ and\
  \citenamefont {Pacheco}}]{van-segbroeck:PRL:2009}%
  \BibitemOpen
  \bibfield  {author} {\bibinfo {author} {\bibfnamefont {S.}~\bibnamefont
  {Van~Segbroeck}}, \bibinfo {author} {\bibfnamefont {F.~C.}\ \bibnamefont
  {Santos}}, \bibinfo {author} {\bibfnamefont {T.}~\bibnamefont {Lenaerts}}, \
  and\ \bibinfo {author} {\bibfnamefont {J.~M.}\ \bibnamefont {Pacheco}},\
  }\href@noop {} {\bibfield  {journal} {\bibinfo  {journal} {Physical Review
  Letters}\ }\textbf {\bibinfo {volume} {102}},\ \bibinfo {pages} {058105}
  (\bibinfo {year} {2009})}\BibitemShut {NoStop}%
\bibitem [{\citenamefont {Black}\ \emph {et~al.}(2012)\citenamefont {Black},
  \citenamefont {Traulsen},\ and\ \citenamefont {Galla}}]{black:PRL:2012}%
  \BibitemOpen
  \bibfield  {author} {\bibinfo {author} {\bibfnamefont {A.~J.}\ \bibnamefont
  {Black}}, \bibinfo {author} {\bibfnamefont {A.}~\bibnamefont {Traulsen}}, \
  and\ \bibinfo {author} {\bibfnamefont {T.}~\bibnamefont {Galla}},\
  }\href@noop {} {\bibfield  {journal} {\bibinfo  {journal} {Physical Review
  Letters}\ }\textbf {\bibinfo {volume} {109}},\ \bibinfo {pages} {028101}
  (\bibinfo {year} {2012})}\BibitemShut {NoStop}%
\bibitem [{\citenamefont {Nowak}\ \emph {et~al.}(2004)\citenamefont {Nowak},
  \citenamefont {Sasaki}, \citenamefont {Taylor},\ and\ \citenamefont
  {Fudenberg}}]{nowak:Nature:2004}%
  \BibitemOpen
  \bibfield  {author} {\bibinfo {author} {\bibfnamefont {M.~A.}\ \bibnamefont
  {Nowak}}, \bibinfo {author} {\bibfnamefont {A.}~\bibnamefont {Sasaki}},
  \bibinfo {author} {\bibfnamefont {C.}~\bibnamefont {Taylor}}, \ and\ \bibinfo
  {author} {\bibfnamefont {D.}~\bibnamefont {Fudenberg}},\ }\href@noop {}
  {\bibfield  {journal} {\bibinfo  {journal} {Nature}\ }\textbf {\bibinfo
  {volume} {428}},\ \bibinfo {pages} {646} (\bibinfo {year}
  {2004})}\BibitemShut {NoStop}%
\bibitem [{\citenamefont {Blume}(1993)}]{blume:GEB:1993}%
  \BibitemOpen
  \bibfield  {author} {\bibinfo {author} {\bibfnamefont {L.~E.}\ \bibnamefont
  {Blume}},\ }\href@noop {} {\bibfield  {journal} {\bibinfo  {journal} {Games
  and Economic Behavior}\ }\textbf {\bibinfo {volume} {5}},\ \bibinfo {pages}
  {387} (\bibinfo {year} {1993})}\BibitemShut {NoStop}%
\bibitem [{\citenamefont {Szab{\'o}}\ and\ \citenamefont {T{\H
  o}ke}(1998)}]{szabo:PRE:1998}%
  \BibitemOpen
  \bibfield  {author} {\bibinfo {author} {\bibfnamefont {G.}~\bibnamefont
  {Szab{\'o}}}\ and\ \bibinfo {author} {\bibfnamefont {C.}~\bibnamefont {T{\H
  o}ke}},\ }\href@noop {} {\bibfield  {journal} {\bibinfo  {journal} {Physical
  Review E}\ }\textbf {\bibinfo {volume} {58}},\ \bibinfo {pages} {69}
  (\bibinfo {year} {1998})}\BibitemShut {NoStop}%
\bibitem [{\citenamefont {B{\"u}rger}(2000)}]{buerger:book:2000}%
  \BibitemOpen
  \bibfield  {author} {\bibinfo {author} {\bibfnamefont {R.}~\bibnamefont
  {B{\"u}rger}},\ }\href@noop {} {\emph {\bibinfo {title} {The Mathematical
  Theory of Selection, Recombination, and Mutation}}}\ (\bibinfo  {publisher}
  {John Wiley and Sons},\ \bibinfo {year} {2000})\BibitemShut {NoStop}%
\bibitem [{\citenamefont {Wu}\ \emph {et~al.}(2013{\natexlab{a}})\citenamefont
  {Wu}, \citenamefont {Gokhale}, \citenamefont {Van~Veelen}, \citenamefont
  {Wang},\ and\ \citenamefont {Traulsen}}]{Wu:EaE:2013}%
  \BibitemOpen
  \bibfield  {author} {\bibinfo {author} {\bibfnamefont {B.}~\bibnamefont
  {Wu}}, \bibinfo {author} {\bibfnamefont {C.~S.}\ \bibnamefont {Gokhale}},
  \bibinfo {author} {\bibfnamefont {M.}~\bibnamefont {Van~Veelen}}, \bibinfo
  {author} {\bibfnamefont {L.}~\bibnamefont {Wang}}, \ and\ \bibinfo {author}
  {\bibfnamefont {A.}~\bibnamefont {Traulsen}},\ }\href@noop {} {\bibfield
  {journal} {\bibinfo  {journal} {Ecology and Evolution}\ }\textbf {\bibinfo
  {volume} {3}},\ \bibinfo {pages} {1276} (\bibinfo {year}
  {2013}{\natexlab{a}})}\BibitemShut {NoStop}%
\bibitem [{\citenamefont {Traulsen}\ \emph {et~al.}(2006)\citenamefont
  {Traulsen}, \citenamefont {Nowak},\ and\ \citenamefont
  {Pacheco}}]{traulsen:PRE:2006b}%
  \BibitemOpen
  \bibfield  {author} {\bibinfo {author} {\bibfnamefont {A.}~\bibnamefont
  {Traulsen}}, \bibinfo {author} {\bibfnamefont {M.~A.}\ \bibnamefont {Nowak}},
  \ and\ \bibinfo {author} {\bibfnamefont {J.~M.}\ \bibnamefont {Pacheco}},\
  }\href@noop {} {\bibfield  {journal} {\bibinfo  {journal} {Physical Review
  E}\ }\textbf {\bibinfo {volume} {74}},\ \bibinfo {pages} {011909} (\bibinfo
  {year} {2006})}\BibitemShut {NoStop}%
\bibitem [{\citenamefont {Karlin}\ and\ \citenamefont
  {Taylor}(1975)}]{karlin:book:1975}%
  \BibitemOpen
  \bibfield  {author} {\bibinfo {author} {\bibfnamefont {S.}~\bibnamefont
  {Karlin}}\ and\ \bibinfo {author} {\bibfnamefont {H.~M.~A.}\ \bibnamefont
  {Taylor}},\ }\href@noop {} {\emph {\bibinfo {title} {A First Course in
  Stochastic Processes}}},\ \bibinfo {edition} {2nd}\ ed.\ (\bibinfo
  {publisher} {Academic},\ \bibinfo {address} {London},\ \bibinfo {year}
  {1975})\BibitemShut {NoStop}%
\bibitem [{\citenamefont {Ewens}(1979)}]{ewens:book:1979}%
  \BibitemOpen
  \bibfield  {author} {\bibinfo {author} {\bibfnamefont {W.~J.}\ \bibnamefont
  {Ewens}},\ }\href@noop {} {\emph {\bibinfo {title} {Mathematical Population
  Genetics}}}\ (\bibinfo  {publisher} {Springer},\ \bibinfo {address}
  {Berlin},\ \bibinfo {year} {1979})\BibitemShut {NoStop}%
\bibitem [{\citenamefont {Fudenberg}\ and\ \citenamefont
  {Imhof}(2006)}]{fudenberg:JET:2006}%
  \BibitemOpen
  \bibfield  {author} {\bibinfo {author} {\bibfnamefont {D.}~\bibnamefont
  {Fudenberg}}\ and\ \bibinfo {author} {\bibfnamefont {L.~A.}\ \bibnamefont
  {Imhof}},\ }\href@noop {} {\bibfield  {journal} {\bibinfo  {journal} {Journal
  of Economic Theory}\ }\textbf {\bibinfo {volume} {131}},\ \bibinfo {pages}
  {251} (\bibinfo {year} {2006})}\BibitemShut {NoStop}%
\bibitem [{\citenamefont {Wu}\ \emph {et~al.}(2012)\citenamefont {Wu},
  \citenamefont {Gokhale}, \citenamefont {Wang},\ and\ \citenamefont
  {Traulsen}}]{wu:JMB:2012}%
  \BibitemOpen
  \bibfield  {author} {\bibinfo {author} {\bibfnamefont {B.}~\bibnamefont
  {Wu}}, \bibinfo {author} {\bibfnamefont {C.~S.}\ \bibnamefont {Gokhale}},
  \bibinfo {author} {\bibfnamefont {L.}~\bibnamefont {Wang}}, \ and\ \bibinfo
  {author} {\bibfnamefont {A.}~\bibnamefont {Traulsen}},\ }\href@noop {}
  {\bibfield  {journal} {\bibinfo  {journal} {Journal of Mathematical Biology}\
  }\textbf {\bibinfo {volume} {64}},\ \bibinfo {pages} {803} (\bibinfo {year}
  {2012})}\BibitemShut {NoStop}%
\bibitem [{\citenamefont {Fu}\ \emph {et~al.}(2008)\citenamefont {Fu},
  \citenamefont {Hauert}, \citenamefont {Nowak},\ and\ \citenamefont
  {Wang}}]{fu:PRE:2008}%
  \BibitemOpen
  \bibfield  {author} {\bibinfo {author} {\bibfnamefont {F.}~\bibnamefont
  {Fu}}, \bibinfo {author} {\bibfnamefont {C.}~\bibnamefont {Hauert}}, \bibinfo
  {author} {\bibfnamefont {M.~A.}\ \bibnamefont {Nowak}}, \ and\ \bibinfo
  {author} {\bibfnamefont {L.}~\bibnamefont {Wang}},\ }\href@noop {} {\bibfield
   {journal} {\bibinfo  {journal} {Physical Review E}\ }\textbf {\bibinfo
  {volume} {78}} (\bibinfo {year} {2008})}\BibitemShut {NoStop}%
\bibitem [{\citenamefont {Sigmund}\ \emph {et~al.}(2010)\citenamefont
  {Sigmund}, \citenamefont {De~Silva}, \citenamefont {Traulsen},\ and\
  \citenamefont {Hauert}}]{sigmund:Nature:2010}%
  \BibitemOpen
  \bibfield  {author} {\bibinfo {author} {\bibfnamefont {K.}~\bibnamefont
  {Sigmund}}, \bibinfo {author} {\bibfnamefont {H.}~\bibnamefont {De~Silva}},
  \bibinfo {author} {\bibfnamefont {A.}~\bibnamefont {Traulsen}}, \ and\
  \bibinfo {author} {\bibfnamefont {C.}~\bibnamefont {Hauert}},\ }\href@noop {}
  {\bibfield  {journal} {\bibinfo  {journal} {Nature}\ }\textbf {\bibinfo
  {volume} {466}},\ \bibinfo {pages} {861} (\bibinfo {year}
  {2010})}\BibitemShut {NoStop}%
\bibitem [{\citenamefont {Lessard}\ and\ \citenamefont
  {Ladret}(2007)}]{lessard:JMB:2007}%
  \BibitemOpen
  \bibfield  {author} {\bibinfo {author} {\bibfnamefont {S.}~\bibnamefont
  {Lessard}}\ and\ \bibinfo {author} {\bibfnamefont {V.}~\bibnamefont
  {Ladret}},\ }\href@noop {} {\bibfield  {journal} {\bibinfo  {journal}
  {Journal of Mathematical Biology}\ }\textbf {\bibinfo {volume} {54}},\
  \bibinfo {pages} {721} (\bibinfo {year} {2007})}\BibitemShut {NoStop}%
\bibitem [{\citenamefont {Lessard}(2011)}]{lessard:DGAA:2011b}%
  \BibitemOpen
  \bibfield  {author} {\bibinfo {author} {\bibfnamefont {S.}~\bibnamefont
  {Lessard}},\ }\href@noop {} {\bibfield  {journal} {\bibinfo  {journal}
  {Dynamic Games and Applications}\ }\textbf {\bibinfo {volume} {1}},\ \bibinfo
  {pages} {408} (\bibinfo {year} {2011})}\BibitemShut {NoStop}%
\bibitem [{\citenamefont {Wu}\ \emph {et~al.}(2013{\natexlab{b}})\citenamefont
  {Wu}, \citenamefont {Garc{\'\i}a}, \citenamefont {Hauert},\ and\
  \citenamefont {Traulsen}}]{wu:PLoSCB:2013}%
  \BibitemOpen
  \bibfield  {author} {\bibinfo {author} {\bibfnamefont {B.}~\bibnamefont
  {Wu}}, \bibinfo {author} {\bibfnamefont {J.}~\bibnamefont {Garc{\'\i}a}},
  \bibinfo {author} {\bibfnamefont {C.}~\bibnamefont {Hauert}}, \ and\ \bibinfo
  {author} {\bibfnamefont {A.}~\bibnamefont {Traulsen}},\ }\href@noop {}
  {\bibfield  {journal} {\bibinfo  {journal} {PLoS Computational Biology}\
  }\textbf {\bibinfo {volume} {9}},\ \bibinfo {pages} {e1003381} (\bibinfo
  {year} {2013}{\natexlab{b}})}\BibitemShut {NoStop}%
\bibitem [{\citenamefont {Hofbauer}\ and\ \citenamefont
  {Sigmund}(1998)}]{hofbauer:book:1998}%
  \BibitemOpen
  \bibfield  {author} {\bibinfo {author} {\bibfnamefont {J.}~\bibnamefont
  {Hofbauer}}\ and\ \bibinfo {author} {\bibfnamefont {K.}~\bibnamefont
  {Sigmund}},\ }\href@noop {} {\emph {\bibinfo {title} {Evolutionary Games and
  Population Dynamics}}}\ (\bibinfo  {publisher} {Cambridge University Press,
  Cambridge},\ \bibinfo {year} {1998})\BibitemShut {NoStop}%
\bibitem [{\citenamefont {Gokhale}\ and\ \citenamefont
  {Traulsen}(2014)}]{gokhale:DGAA:2014}%
  \BibitemOpen
  \bibfield  {author} {\bibinfo {author} {\bibfnamefont {C.~S.}\ \bibnamefont
  {Gokhale}}\ and\ \bibinfo {author} {\bibfnamefont {A.}~\bibnamefont
  {Traulsen}},\ }\href@noop {} {\bibfield  {journal} {\bibinfo  {journal}
  {Dynamic Games and Applications}\ } (\bibinfo {year} {2014})}\BibitemShut
  {NoStop}%
\bibitem [{\citenamefont {Pe{\~n}a}\ \emph {et~al.}(2014)\citenamefont
  {Pe{\~n}a}, \citenamefont {Lehmann},\ and\ \citenamefont
  {N{\"o}ldeke}}]{pena:JTB:2014}%
  \BibitemOpen
  \bibfield  {author} {\bibinfo {author} {\bibfnamefont {J.}~\bibnamefont
  {Pe{\~n}a}}, \bibinfo {author} {\bibfnamefont {L.}~\bibnamefont {Lehmann}}, \
  and\ \bibinfo {author} {\bibfnamefont {G.}~\bibnamefont {N{\"o}ldeke}},\
  }\href@noop {} {\bibfield  {journal} {\bibinfo  {journal} {Journal of
  Theoretical Biology}\ }\textbf {\bibinfo {volume} {346}},\ \bibinfo {pages}
  {23} (\bibinfo {year} {2014})}\BibitemShut {NoStop}%
\bibitem [{\citenamefont {Wu}\ \emph {et~al.}(2010)\citenamefont {Wu},
  \citenamefont {Altrock}, \citenamefont {Wang},\ and\ \citenamefont
  {Traulsen}}]{wu:PRE:2010}%
  \BibitemOpen
  \bibfield  {author} {\bibinfo {author} {\bibfnamefont {B.}~\bibnamefont
  {Wu}}, \bibinfo {author} {\bibfnamefont {P.~M.}\ \bibnamefont {Altrock}},
  \bibinfo {author} {\bibfnamefont {L.}~\bibnamefont {Wang}}, \ and\ \bibinfo
  {author} {\bibfnamefont {A.}~\bibnamefont {Traulsen}},\ }\href@noop {}
  {\bibfield  {journal} {\bibinfo  {journal} {Physical Review E}\ }\textbf
  {\bibinfo {volume} {82}},\ \bibinfo {pages} {046106} (\bibinfo {year}
  {2010})}\BibitemShut {NoStop}%
\bibitem [{\citenamefont {Pinheiro}\ \emph {et~al.}(2012)\citenamefont
  {Pinheiro}, \citenamefont {Santos},\ and\ \citenamefont
  {Pacheco}}]{pinheiro:NJP:2012}%
  \BibitemOpen
  \bibfield  {author} {\bibinfo {author} {\bibfnamefont {F.~L.}\ \bibnamefont
  {Pinheiro}}, \bibinfo {author} {\bibfnamefont {F.~C.}\ \bibnamefont
  {Santos}}, \ and\ \bibinfo {author} {\bibfnamefont {J.~M.}\ \bibnamefont
  {Pacheco}},\ }\href@noop {} {\bibfield  {journal} {\bibinfo  {journal} {New
  Journal of Physics}\ }\textbf {\bibinfo {volume} {14}},\ \bibinfo {pages}
  {073035} (\bibinfo {year} {2012})}\BibitemShut {NoStop}%
\bibitem [{\citenamefont {Traulsen}\ \emph {et~al.}(2012)\citenamefont
  {Traulsen}, \citenamefont {Claussen},\ and\ \citenamefont
  {Hauert}}]{traulsen:PRE:2012}%
  \BibitemOpen
  \bibfield  {author} {\bibinfo {author} {\bibfnamefont {A.}~\bibnamefont
  {Traulsen}}, \bibinfo {author} {\bibfnamefont {J.~C.}\ \bibnamefont
  {Claussen}}, \ and\ \bibinfo {author} {\bibfnamefont {C.}~\bibnamefont
  {Hauert}},\ }\href@noop {} {\bibfield  {journal} {\bibinfo  {journal}
  {Physical Review E}\ }\textbf {\bibinfo {volume} {85}},\ \bibinfo {pages}
  {041901} (\bibinfo {year} {2012})}\BibitemShut {NoStop}%
\bibitem [{\citenamefont {Bladon}\ \emph {et~al.}(2010)\citenamefont {Bladon},
  \citenamefont {Galla},\ and\ \citenamefont {McKane}}]{bladon:PRE:2010}%
  \BibitemOpen
  \bibfield  {author} {\bibinfo {author} {\bibfnamefont {A.~J.}\ \bibnamefont
  {Bladon}}, \bibinfo {author} {\bibfnamefont {T.}~\bibnamefont {Galla}}, \
  and\ \bibinfo {author} {\bibfnamefont {A.~J.}\ \bibnamefont {McKane}},\
  }\href@noop {} {\bibfield  {journal} {\bibinfo  {journal} {Physical Review
  E}\ }\textbf {\bibinfo {volume} {81}},\ \bibinfo {pages} {066122} (\bibinfo
  {year} {2010})}\BibitemShut {NoStop}%
\end{thebibliography}
\end{document}